\documentclass[sigconf,edbt]{acmart-edbt2019}

\usepackage{booktabs} 
\usepackage{graphicx}

\usepackage[export]{adjustbox}
\usepackage{multirow}
\usepackage{subfig}
\usepackage{amssymb}
\usepackage{wrapfig}
\usepackage{amsmath}
\usepackage{epstopdf}
\usepackage{bigstrut}
\usepackage{booktabs}
\usepackage{caption}
\usepackage[misc]{ifsym}

\usepackage{tikz}
\usetikzlibrary{trees}
\usetikzlibrary{calc}
\usetikzlibrary{arrows}
\tikzstyle{vertex}=[circle, draw]
\usepackage{pgfplots}
\usetikzlibrary{shapes}
\usetikzlibrary{shapes.geometric}
\usetikzlibrary{arrows,decorations.pathmorphing,backgrounds,positioning,fit}
\usetikzlibrary{arrows,shapes,positioning,shadows,trees}
\usetikzlibrary{shapes,shadows,trees}

\makeatletter \pgfkeys{/pgf/.cd,
  parallelepiped offset x/.initial=2mm,
  parallelepiped offset y/.initial=2mm
} \pgfdeclareshape{parallelepiped} {
  \inheritsavedanchors[from=rectangle] 
  \inheritanchorborder[from=rectangle]
  \inheritanchor[from=rectangle]{north}
  \inheritanchor[from=rectangle]{north west}
  \inheritanchor[from=rectangle]{north east}
  \inheritanchor[from=rectangle]{center}
  \inheritanchor[from=rectangle]{west}
  \inheritanchor[from=rectangle]{east}
  \inheritanchor[from=rectangle]{mid}
  \inheritanchor[from=rectangle]{mid west}
  \inheritanchor[from=rectangle]{mid east}
  \inheritanchor[from=rectangle]{base}
  \inheritanchor[from=rectangle]{base west}
  \inheritanchor[from=rectangle]{base east}
  \inheritanchor[from=rectangle]{south}
  \inheritanchor[from=rectangle]{south west}
  \inheritanchor[from=rectangle]{south east}
  \backgroundpath{
    \southwest \pgf@xa=\pgf@x \pgf@ya=\pgf@y
    \northeast \pgf@xb=\pgf@x \pgf@yb=\pgf@y
    \pgfmathsetlength\pgfutil@tempdima{\pgfkeysvalueof{/pgf/parallelepiped offset x}}
    \pgfmathsetlength\pgfutil@tempdimb{\pgfkeysvalueof{/pgf/parallelepiped offset y}}
    \def\ppd@offset{\pgfpoint{\pgfutil@tempdima}{\pgfutil@tempdimb}}
    \pgfpathmoveto{\pgfqpoint{\pgf@xa}{\pgf@ya}}
    \pgfpathlineto{\pgfqpoint{\pgf@xb}{\pgf@ya}}
    \pgfpathlineto{\pgfqpoint{\pgf@xb}{\pgf@yb}}
    \pgfpathlineto{\pgfqpoint{\pgf@xa}{\pgf@yb}}
    \pgfpathclose
    \pgfpathmoveto{\pgfqpoint{\pgf@xb}{\pgf@ya}}
    \pgfpathlineto{\pgfpointadd{\pgfpoint{\pgf@xb}{\pgf@ya}}{\ppd@offset}}
    \pgfpathlineto{\pgfpointadd{\pgfpoint{\pgf@xb}{\pgf@yb}}{\ppd@offset}}
    \pgfpathlineto{\pgfpointadd{\pgfpoint{\pgf@xa}{\pgf@yb}}{\ppd@offset}}
    \pgfpathlineto{\pgfqpoint{\pgf@xa}{\pgf@yb}}
    \pgfpathmoveto{\pgfqpoint{\pgf@xb}{\pgf@yb}}
    \pgfpathlineto{\pgfpointadd{\pgfpoint{\pgf@xb}{\pgf@yb}}{\ppd@offset}}
  }
}

\setcopyright{rightsretained}

\acmDOI{}

\acmISBN{978-3-89318-081-3.}

\acmConference[EDBT 2019]{22nd International Conference on Extending Database Technology (EDBT)}{March 26-29, 2019}{Lisbon, Portugal} 
\acmYear{2019}

\begin{document}
	\title{SCube: A Tool for Segregation Discovery}
\author{Alessandro Baroni}
\affiliation{%
	\institution{University of Pisa, Italy}
	\streetaddress{Largo B. Pontecorvo 3}
	\postcode{56127}
}
\email{baroni@di.unipi.it}

\author{Salvatore Ruggieri}
\affiliation{%
	\institution{University of Pisa, Italy}
	\streetaddress{Largo B. Pontecorvo 3}
	\postcode{56127}
}
\email{ruggieri@di.unipi.it}

\renewcommand{\shortauthors}{}

\begin{abstract}
Segregation is the separation of social groups in the physical or in the online world. Segregation discovery consists of finding contexts of segregation. 
In the modern digital society, discovering segregation is challenging, due to the large amount and the variety of social data. We present a tool in support of segregation discovery from relational and graph data. The SCube system builds on attributed graph clustering and frequent itemset mining. It offers to the analyst a multi-dimensional segregation data cube for exploratory data analysis. The demonstration first guides the audience through the relevant social science concepts. Then, it focuses on scenarios around case studies of gender occupational segregation. Two real and large datasets about the boards of directors of Italian and Estonian companies will be explored in search of segregation contexts. The architecture of the SCube system and its computational efficiency challenges and solutions are discussed. 
%
\end{abstract}

\maketitle

\section{Social Segregation}

Ethical issues in data and knowledge management are gaining momentum in the last few years. 
In addition to the traditional field of privacy, techniques for data analysis are being designed or enhanced to take into account moral values such as fairness, transparency, accountability, and diversity\footnote{See e.g., the \textit{Toronto declaration} at \href{https://www.accessnow.org/toronto-declaration}{www.accessnow.org/toronto-declaration}.}. 
We have recentely developed a novel data-driven technique for addressing segregation of social groups through multi-dimen\-sional data analysis \cite{Baroni2017jiis}. The approach is implemented in the SCube system, which we propose to demonstrate using real case studies.

\emph{Social segregation} refers to the ``separation of socially defined groups''~\cite{massey2016segregation}. People are partitioned into
two or more groups on the grounds of personal or cultural traits that can foster discrimination, such as gender, age, ethnicity, income, skin
color, language, religion, political opinion, membership to a national minority, etc. 
Contact, communication, or interaction among groups are limited
by their physical, working or socio-economic distance. This can be observed when dissecting society in organizational units (neighborhoods, schools,
job types).
Due to the ubiquitous presence and pervasiveness of ICT, segregation is shifting from ancient forms of well explored spatial segregation\footnote{See census stats, e.g.,~\href{https://www.census.gov/topics/housing/housing-patterns/data.html}{www.census.gov/topics/housing/housing-patterns/data.html}} to novel forms of digital segregation. For instance, it has been warned that the filter bubble generated by personalization of online social networks may foster idelogical
segregation~\cite{flaxman2013ideological}, opinion polarization~\cite{maes2015will}, and informational segregation. 
A data-driven technology that enables the assessment of the extent, nature, and trends of social segregation in the offline or online world, is of extreme interest for a wide audience: social scientistics, public policy makers, regulation and control authorities, professional associations, civil rights societies, and investigative journalists. Business decision makers should also care of business practices, particularly automated decision making, that segregate customers and products through stereotypes, because this limits diversity and reduces opportunities of cross-selling. 
Finally, data scientists and professionals should be aware of the unintended consequences of their models (recommender systems, link suggestion systems, classifiers) on the cohesion of society at large.

\begin{figure}[t]
	\mbox{}\\[2ex]
	\centering
	\noindent 
	\begin{minipage}[b]{0.50\textwidth}\centering
		\scalebox{0.46}{
			\begin{tikzpicture}
			\node at (0.05, -0.65)  {\textbf{young}};
			\node at (1.65, -0.65)  {\textbf{middle}};
			\node at (2.5, -1.1)  {\textbf{age (SA)}};
			\node at (3.3, -0.65)  {\textbf{elder}};
			\node at (4.95, -0.65)  {\textbf{$\ast$}};
			\node  at (0.6, 6) [rotate=55] {\textbf{region (CA)}};
			
			\node at (-1.42, 0.05)  {\textbf{female}};
			\node at (-2.8, 0.85)  {\textbf{sex (SA)}};
			\node at (-1.32, 0.85)  {\textbf{male}};
			\node at (-1.12, 1.7)  {\textbf{$\ast$}};
			
			\draw (-0.77,2.05)  -- (3.27,8.09); 
			\node at (3.1, 8.3)  {\textbf{$\ast$}};
			\node at (-0.85,2.55)  {\textbf{north}};
			\node at (1.1,5.55)  {\textbf{south}};
			\draw (-0.77, -0.34)  -- (3.27,5.7); 
			\node[parallelepiped,draw=black,fill=white,
			minimum width=1.5cm,minimum height=0.7cm] (1) at (0,0)  {\textbf{0.78}} ;  
			\node[parallelepiped,draw=black,fill=white,
			minimum width=1.5cm,minimum height=0.7cm] (2)  at (1.65,0) {\textbf{0.63}};
			\node[parallelepiped,draw=black,fill=white,
			minimum width=1.5cm,minimum height=0.7cm] (3)  at (3.3,0) {\textbf{-}};
			\node[parallelepiped,draw=black,fill=gray!20,
			minimum width=1.5cm,minimum height=0.7cm] (3)  at (4.95,0) {\textbf{0.71}};
			\node[parallelepiped,draw=black,fill=white,
			minimum width=1.5cm,minimum height=0.7cm] (1) at (0,0.85) {\textbf{0.71}} ; 
			\node[parallelepiped,draw=black,fill=white,
			minimum width=1.5cm,minimum height=0.7cm] (2)  at (1.65,0.85) {\textbf{0.63}};
			\node[parallelepiped,draw=black,fill=white,
			minimum width=1.5cm,minimum height=0.7cm] (3)  at (3.3,0.85) {\textbf{0.88}};
			\node[parallelepiped,draw=black,fill=gray!20,
			minimum width=1.5cm,minimum height=0.7cm] (3)  at (4.95,0.85) {\textbf{0.71}};
			\node[parallelepiped,draw=black,fill=gray!20,
			minimum width=1.5cm,minimum height=0.7cm] (1) at (0,1.70) {\textbf{0.50}} ; 
			
			\node[parallelepiped,draw=black,fill=gray!20,
			minimum width=1.5cm,minimum height=0.7cm] (2)  at (1.65,1.70) {\textbf{0.83}};
			
			\node[parallelepiped,draw=black,fill=gray!20,
			minimum width=1.5cm,minimum height=0.7cm] (3)  at (3.3,1.70) {\textbf{-}};
			
			\node[parallelepiped,draw=black,fill=gray!40,
			minimum width=1.5cm,minimum height=0.7cm] (3)  at (4.95,1.70) {\textbf{-}};
			
			\node[parallelepiped,draw=black,fill=white,
			minimum width=1.5cm,minimum height=0.7cm] (1) at (2,3) {\textbf{-}} ; 
			\node[parallelepiped,draw=black,fill=white,
			minimum width=1.5cm,minimum height=0.7cm] (2)  at (3.65,3) {\textbf{0.43}};
			\node[parallelepiped,draw=black,fill=white,
			minimum width=1.5cm,minimum height=0.7cm] (3)  at (5.3,3) {\textbf{0.86}};
			\node[parallelepiped,draw=black,fill=gray!20,
			minimum width=1.5cm,minimum height=0.7cm] (3)  at (6.95,3) {\textbf{0.75}};
			\node[parallelepiped,draw=black,fill=white,
			minimum width=1.5cm,minimum height=0.7cm] (1) at (2,3.85) {\textbf{0.75}} ; 
			\node[parallelepiped,draw=black,fill=white,
			minimum width=1.5cm,minimum height=0.7cm] (2)  at (3.65,3.85) {\textbf{0.50}};
			\node[parallelepiped,draw=black,fill=white,
			minimum width=1.5cm,minimum height=0.7cm] (3)  at (5.3,3.85) {\textbf{0.88}};
			\node[parallelepiped,draw=black,fill=gray!20,
			minimum width=1.5cm,minimum height=0.7cm] (3)  at (6.95,3.85) {\textbf{0.75}};
			\node[parallelepiped,draw=black,fill=gray!20,
			minimum width=1.5cm,minimum height=0.7cm] (1) at (2,4.70) {\textbf{-}} ; 
			\node[parallelepiped,draw=black,fill=gray!20,
			minimum width=1.5cm,minimum height=0.7cm] (2)  at (3.65,4.70) {\textbf{0.35}};
			\node[parallelepiped,draw=black,fill=gray!20,
			minimum width=1.5cm,minimum height=0.7cm] (3)  at (5.3,4.70) {\textbf{0.67}};
			\node[parallelepiped,draw=black,fill=gray!40,
			minimum width=1.5cm,minimum height=0.7cm] (3)  at (6.95,4.70) {\textbf{-}};
			\node[parallelepiped,draw=black,fill=gray!20,
			minimum width=1.5cm,minimum height=0.7cm] (1) at (4,6) {\textbf{0.83}} ; 
			\node[parallelepiped,draw=black,fill=gray!20,
			minimum width=1.5cm,minimum height=0.7cm] (2)  at (5.65,6) {\textbf{0.22}};
			\node[parallelepiped,draw=black,fill=gray!20,
			minimum width=1.5cm,minimum height=0.7cm] (3)  at (7.3,6) {\textbf{0.76}};
			\node[parallelepiped,draw=black,fill=gray!40,
			minimum width=1.5cm,minimum height=0.7cm] (3)  at (8.95,6) {\textbf{0.30}};
			\node[parallelepiped,draw=black,fill=gray!20,
			minimum width=1.5cm,minimum height=0.7cm] (1) at (4,6.85) {\textbf{0.62}} ; 
			\node[parallelepiped,draw=black,fill=gray!20,
			minimum width=1.5cm,minimum height=0.7cm] (2)  at (5.65,6.85) {\textbf{0.57}};
			\node[parallelepiped,draw=black,fill=gray!20,
			minimum width=1.5cm,minimum height=0.7cm] (3)  at (7.3,6.85) {\textbf{0.56}};
			\node[parallelepiped,draw=black,fill=gray!40,
			minimum width=1.5cm,minimum height=0.7cm] (3)  at (8.95,6.85) {\textbf{0.30}};
			\node[parallelepiped,draw=black,fill=gray!40,
			minimum width=1.5cm,minimum height=0.7cm] (1) at (4,7.70) {\textbf{0.46}} ; 
			\node[parallelepiped,draw=black,fill=gray!40,
			minimum width=1.5cm,minimum height=0.7cm] (2)  at (5.65,7.70) {\textbf{0.59}};
			\node[parallelepiped,draw=black,fill=gray!40,
			minimum width=1.5cm,minimum height=0.7cm] (3)  at (7.3,7.70) {\textbf{0.66}};
			\node[parallelepiped,draw=black,fill=gray!60,
			minimum width=1.5cm,minimum height=0.7cm] (3)  at (8.95,7.70) {\textbf{-}};
			\draw (5.9,2.25)  -- (9.92,8.25); 
			\draw (5.9,-0.17)  -- (9.925,5.86); 
			\end{tikzpicture}
		}
	\end{minipage}
	\caption{A segregation data cube with dissimilarity index.\label{fig:DataCube}}
	\vspace{-5mm}
\end{figure}
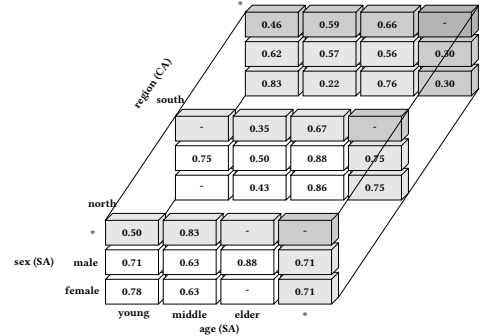

\section{Segregation Discovery}

From a data analysis perspective, the key problem of assessing social segregation has been investigated so far by hypothesis testing, i.e.,~by formulating one or more possible contexts of segregation against a certain social group, and then in empirically testing such hypotheses. Such an approach is currently supported by statistical tools, such as the R packages \textit{OasisR}\footnote{\href{https://cran.r-project.org/package=OasisR}{cran.r-project.org/package=OasisR}} and \textit{seg}\footnote{\href{https://cran.r-project.org/package=seg}{cran.r-project.org/package=seg}} \cite{seg2014}, or by GIS tools such as the  \textit{Geo-Segregation Analyzer}\footnote{\href{http://geoseganalyzer.ucs.inrs.ca}{geoseganalyzer.ucs.inrs.ca}} \cite{geosa2014}.
%
%
%
The formulation of an hypothesis, however, is not straightforward, and it is potentially biased by the expectations of the data analyst of finding segregation in a certain context. In addition, exploration of multiple hypothesis can be time consuming, since data have to be processed multiple times. Finally, this approach is subject to erroneous conclusions if data is considered at wrong granularity -- an instance of the Simpson's paradox.

\textbf{Multi-dimensional segregation data cube.}
Our approach consists of providing the analysts with a multi-dimensional data cube that can be explored in search of candidate contexts of segregation. An example segregation data cube is shown in  Fig.\ref{fig:DataCube}. Dimensions of the data cube include two types of attributes:
\begin{itemize}
	\item \emph{segregation
		attributes} (SA), such as \texttt{\small sex}, \texttt{\small age}, and
	\texttt{\small ethni\-city}, which denote (minority/protected) groups potentially exposed to
	segregation;
	\item \emph{context attributes} (CA), such as
	\texttt{\small region} and \texttt{\small job type}, which denote contexts where
	segregation may appear. 
\end{itemize}
Metrics of the data cube are chosen among the social science indexes proposed for measuring the degree of segregation of social groups within a society \cite{massey1988dimensions}. Here, we recall only one such index, but the SCube system is parametric to the indexes and it computes 6 of them: dissimilarity, Gini, Information index, Isolation, Interaction, Atkinson. Also, we restrict to binary groups (minority/majority). 
Let $T$ be the size of the total population under consideration, $0 < M < T$ be the size of a
minority group, $T-M$ the size of the rest of society  
(or majority group) and $P = M/T$ be the overall fraction of the minority group. Assume that there are $n$ organizational units (or simply, units -- such as schools, neighboorhoods, job types, etc.), and that for $i \in [1, n]$, $t_i$ is the size of the population in unit $i$, and $m_i$ is the size of the minority group in unit $i$. 
The \emph{dissimilarity index} $D$ measures the absolute distance between the fractions of minority and majority groups over the units:
\setlength\abovedisplayskip{0pt}
\setlength\belowdisplayskip{0pt}
\begin{equation*}
D = \frac{1}{2} \sum_{i=1}^n \left|\frac{m_i}{M} - \frac{t_i -
	m_i}{T-M}\right| \label{equ:dissimilarity2}
\end{equation*}
$D$ ranges over $[0, 1]$, with higher values denoting higher segregation. Dissimilarity is minimum when for all $i \in [1, n]$,
$m_i/t_i = M/T$, namely the distribution of the minority group is
uniform over units. It is maximum when for all $i \in [1, n]$,
either $m_i = t_i$ or $m_i = 0$, namely every unit includes members
of only one group (complete segregation).
Dissimilarity and other segregation indexes can be interpreted as metrics in a cell of a multi-dimensional cube as follows: set the total population as those individuals that satisfy the CA coordinates of the cell; and, set the minority population as those individuals that satisfy the SA coordinates. For instance, the cube cell in Fig.\ref{fig:DataCube} with SA coordinates $\texttt{\small sex=female, age=young}$ and CA coordinates $\texttt{\small region=north}$ contains the dissimilarity index for the population living in the north region and for the minority group of young women. Notice that the number $n$ of organizational units here have to be determined \textit{a-priori}, while the total population and minority groups in each unit depend on the values of cell coordinates. As in standard multi-dimensional modelling \cite{DBLP:journals/datamine/GrayCBLRVPP97}, the special value ``$\star$" allows for considering different granularities of analysis.

\textbf{Segregation analysis of tabular data.}
We assume in input a relational table with a tuple for every individual in the population, including SA and CA attributes, and with a further attribute \texttt{\small unitID} which denotes the unit an individual belongs to. 
Unfortunately, segregation indexes are not additive metrics (see \cite{Baroni2017jiis}). This gives rise to the problem of efficiently computing a data cube for segregation analysis. 
Our approach is more specialized than generic holistic aggregate computation in datacubes \cite{DBLP:journals/tkde/NandiYBR12}. We resort to frequent closed itemset mining \cite{FPM2007}.
Data cube coordinates are encoded into itemsets of the form $\mathbf{A}, \mathbf{B}$, where $\mathbf{A}$ denotes a minority subgroup  and $\mathbf{B}$ denotes a context.
Recalling the previous example,  $\mathbf{A}=$\texttt{\small sex=female, age=young} defines the SA coordinates, and
$\mathbf{B}=$\texttt{\small region=north} defines the CA coordinates. The \textit{SegregationDataCubeBuilder} algorithm described in \cite{Baroni2017jiis} fills data cube cells with the value of a segregation index by scanning frequent closed itemsets  of the form above. 
Since relational data is transformed into transaction data\-ba\-se for itemset mining, we obtain for free that CA or SA attributes can be multi-valued, e.g.,~to denote that an individual owns both a house and a car we admit a relation tuple $\sigma$ such that $\sigma[\texttt{\small owns}] =  \{\texttt{\small house}, \texttt{\small car}\}$.

\begin{figure}[t]
	\centering
	\includegraphics[trim=0pt 0pt 0pt 10pt,width=.36\textwidth]{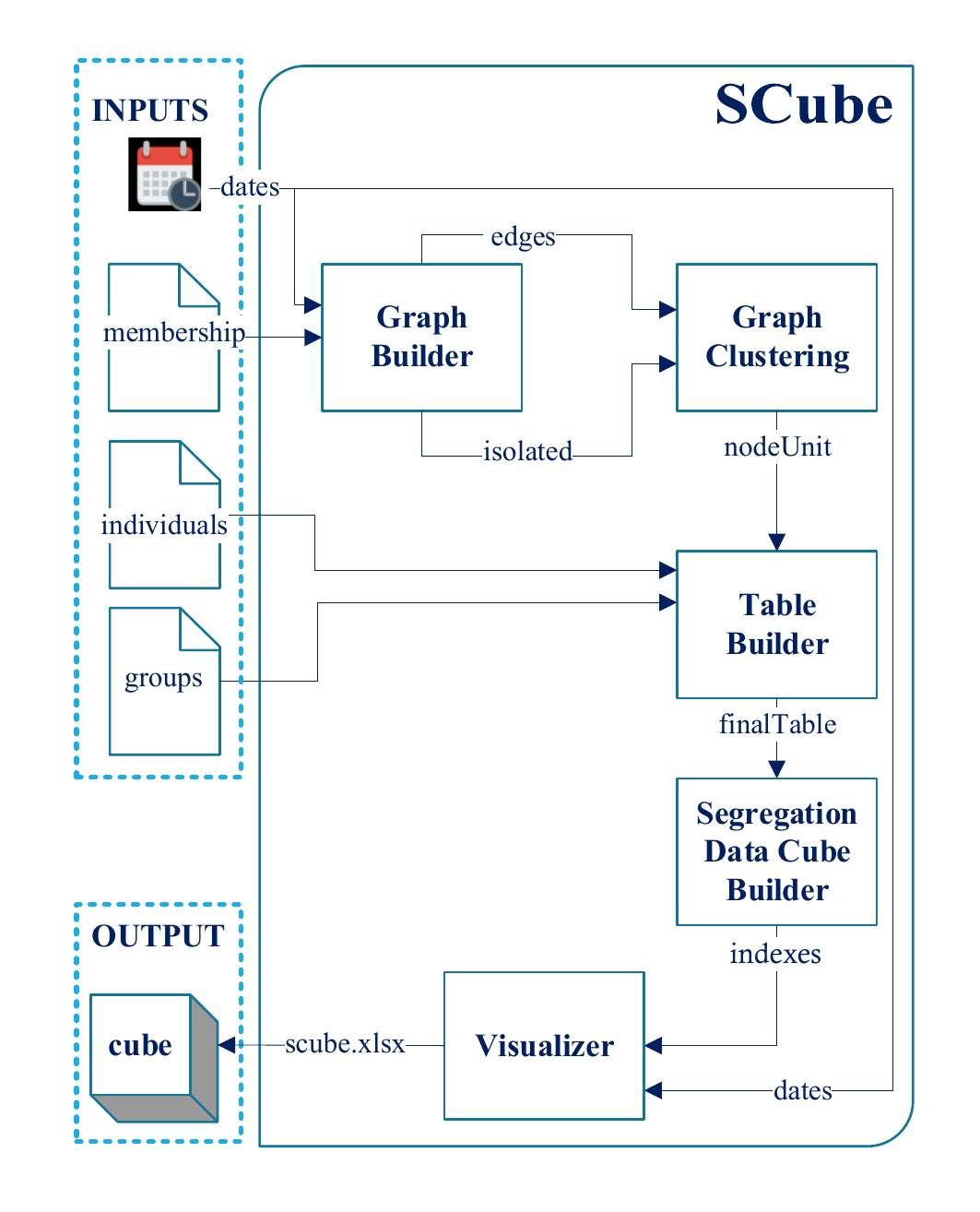}
	\vspace{-5mm}
	\caption{SCube architecture.\label{fig:arch}}
	\vspace{-5mm}
\end{figure}

\begin{figure*}[t]
	\begin{minipage}{0.76\textwidth}
	    \hspace{-3ex}
		\includegraphics[scale=0.63,valign=t]{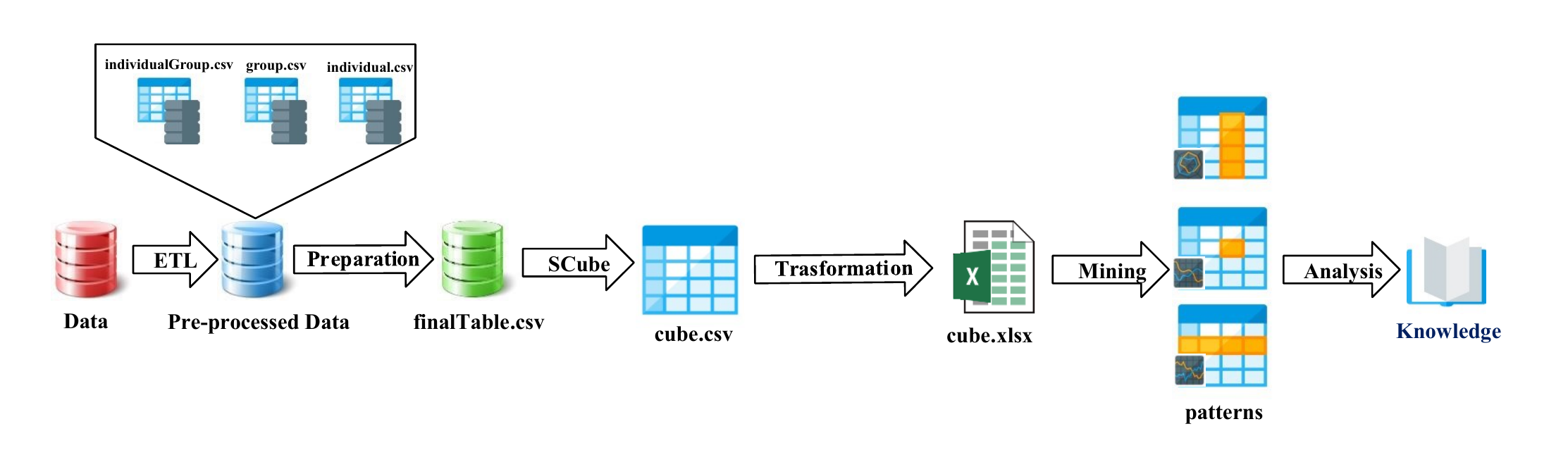}
		\small
		\begin{tabular}{|c|c|c|c|c|c|}	
			\hline
			\multicolumn{3}{|c|}{\emph{\textbf{segregation atts}}} & \multicolumn{2}{|c|}{\emph{\textbf{context atts}}} & \\
			\hline
			\textbf{gender} & \textbf{age} & \textbf{birthplace} & \textbf{residence} & \textbf{sector} & \textbf{unitID} \\
			\hline\hline
			\texttt{M} & \texttt{15-38} & \texttt{foreign} & \texttt{north} & $\{$\texttt{education}$\}$ & 1\\
			\texttt{F} & \texttt{39-46} & \texttt{south} & \texttt{south} & $\{$\texttt{electricity}, \texttt{transports}$\}$ & 2\\
			\texttt{M} & \texttt{55-65} & \texttt{north} & \texttt{south} & $\{$\texttt{agriculture}$\}$ & 1\\
			\ldots & \ldots & \ldots & \ldots & \ldots & \ldots \\
			\hline
		\end{tabular}
	\end{minipage}\hfill
	\begin{minipage}{0.22\textwidth} \hspace{5ex}
		\includegraphics[width=0.9\textwidth,valign=t]{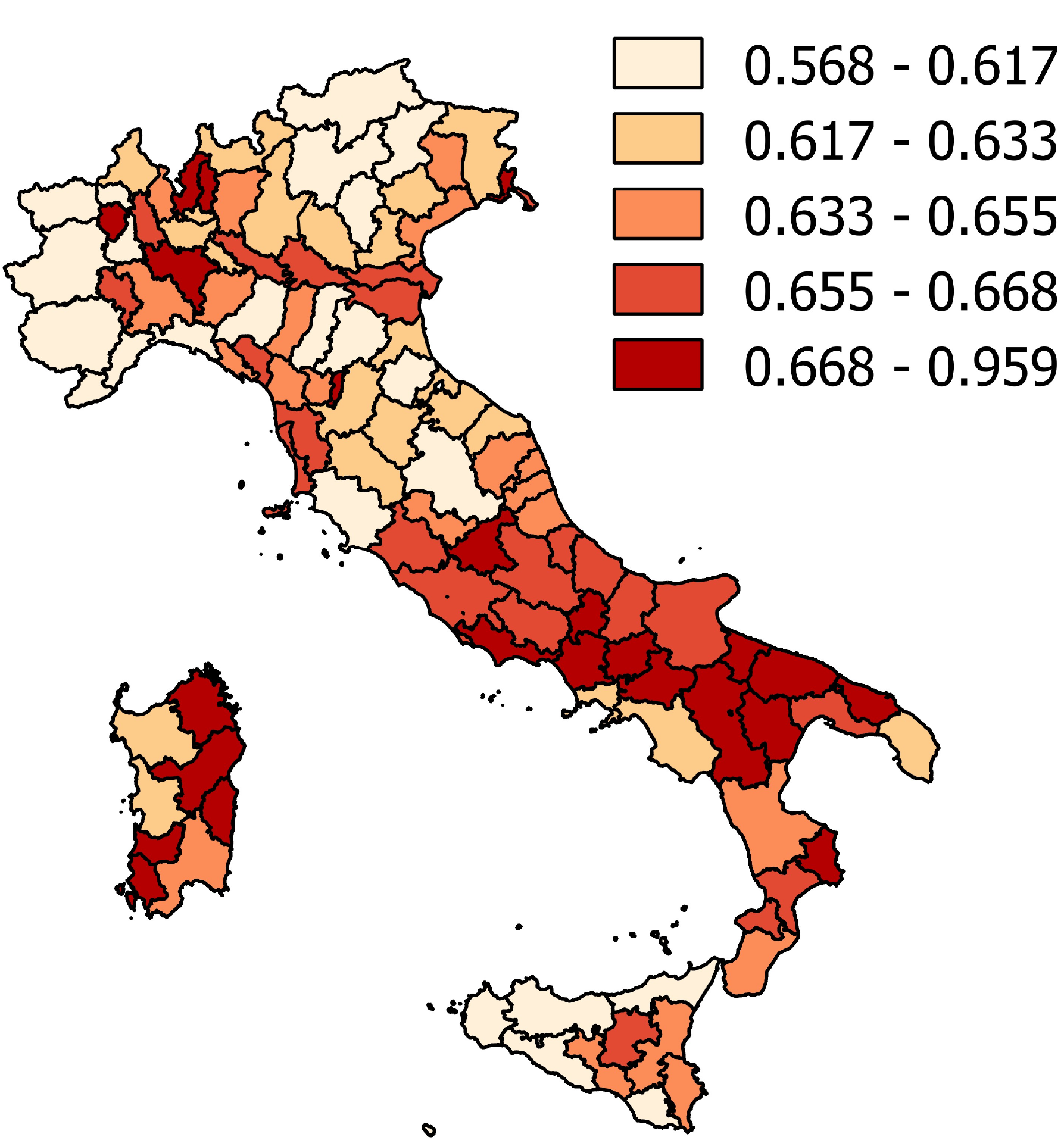}
	\end{minipage}
	\caption{The process of segregation discovery supported by SCube (left, top), input to \textbf{SegregationDataCubeBuilder} (left, bottom), and an output report on dissimilarity segregation index of the Italian provinces (right).}
	\label{fig:KDDProcess}
	\vspace{-2mm}
\end{figure*}

\textbf{Segregation analysis of graph data.}
While transaction data\-bases are able to cover typical analysis from traditional social science, they are not enough powerful to deal with social network data. We formalize such a case using attributed graphs, where nodes are assigned values on a specified set
of attributes. However, in this scenario, there is no \textit{a-priori} defined
notion of organizational unit, i.e.,~the \texttt{\small unitID} attribute assumed in input so far. Some forms of community discovery using graph clustering become necessary in order to determine the organizational units. 
 Clustering attributed graphs consists of
partitioning them into disjoint communities of nodes that are both
well connected and similar with respect to their attributes \cite{bothorel2015clustering}.
In summary, attributed graph clustering can be used first to partition a social network into communities. At this stage, every node/individual in a community is described by its attributes and the community id, which will be our \texttt{\small unitID} attribute. We have thus reduced the problem to the analysis of relational data, for which the \textit{SegregationDataCubeBuilder} algorithm can be applied. 

\textbf{Segregation analysis of bipartite graphs.}
An even more complex scenario is when individuals are not connected among them, e.g.,~because they are friends, but through a connection with another entity, e.g.,~because they work in the same company. Here, a form of projection on unipartite graph is needed to reduce to the previous case. For instance, in \cite{Baroni2017jiis}, we adopt a bipartite projection 
of the bipartite graph of directors and companies to obtain a graph of companies connected by shared directors. Using projection, we have reduced the problem to the previous case, where attributed graph clustering can be adopted to find communities of companies, which then represent the organizational units for segregation analysis. 

\section{SCube Architecture}

The architecture of SCube is shown in Fig.~\ref{fig:arch}. The system is developed in Java, and it relies on a few state-of-the-art libraries\footnote{\small EWAH for
compressed bitmaps
(\href{https://github.com/lemire/javaewah/}{github.com/lemire/javaewah}),
Apache POI for OOXML docs
(\href{https://poi.apache.org}{poi.apache.org}), Borgelt's
FPGrowth for frequent itemset mining
(\href{http://www.borgelt.net/fpgrowth.html}{www.borgelt.net}),
FastUtil for graph storage
(\href{http://fastutil.di.unimi.it/}{fastutil.di.unimi.it}). }.

\textbf{Inputs.} The user has to provide features for two
entities: \emph{individuals} and \emph{groups}. In the reference case studies, individuals are directors and groups are companies. The input
\texttt{\small individuals} (a CSV file or a JDBC query)  provides for
each individual an ID and a number of attribute values, distinguished
into segregation attributes (e.g.,~\texttt{\small gender}, \texttt{\small age}, \texttt{\small birthplace}) and context attributes (e.g.,~\texttt{\small residence}). A second input
\texttt{\small groups} provides for each group an ID and a number of
context attributes values (e.g.,~industrial \texttt{\small sector} of a
company and its headquarter \texttt{\small location}).
Notice that individuals are subjects to possible segregation, while groups are not.
For this reasons, groups have no SA feature.
A third input is
\texttt{\small membership}, which includes the edges of the bipartite graph of individuals and groups, i.e.,~all pairs (\texttt{\small individualID}, \texttt{\small groupID}) for which the individual is 
related to the group. In our case studies, directors are related to
companies they sit in the board of. We also admit that the pairs
are labeled with a time interval of validity, thus allowing for 
temporal analysis of segregation. We have such an information for the
Estonian dataset. 
A fourth input is a list of snapshot dates at which to consider snapshots of the membership relation.


\begin{figure*}[t]
	\centering
	\includegraphics[width=0.33\textwidth,valign=t]{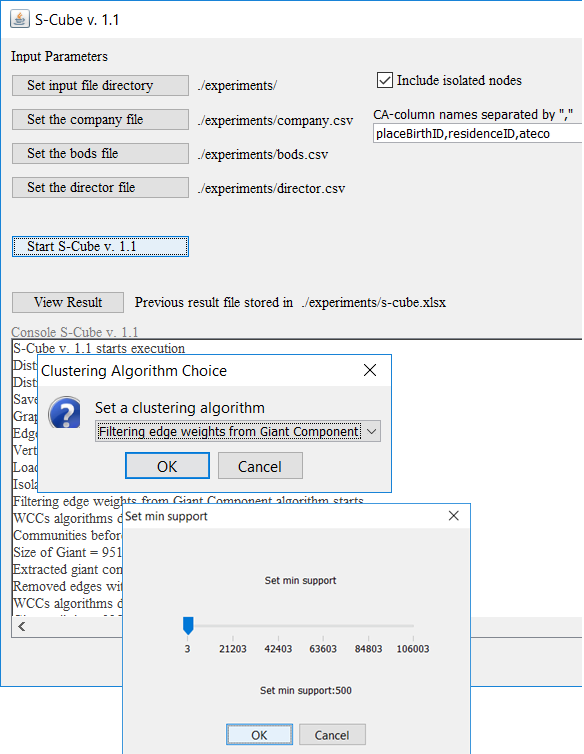}\hspace{3ex}
	\includegraphics[width=0.63\textwidth,valign=t]{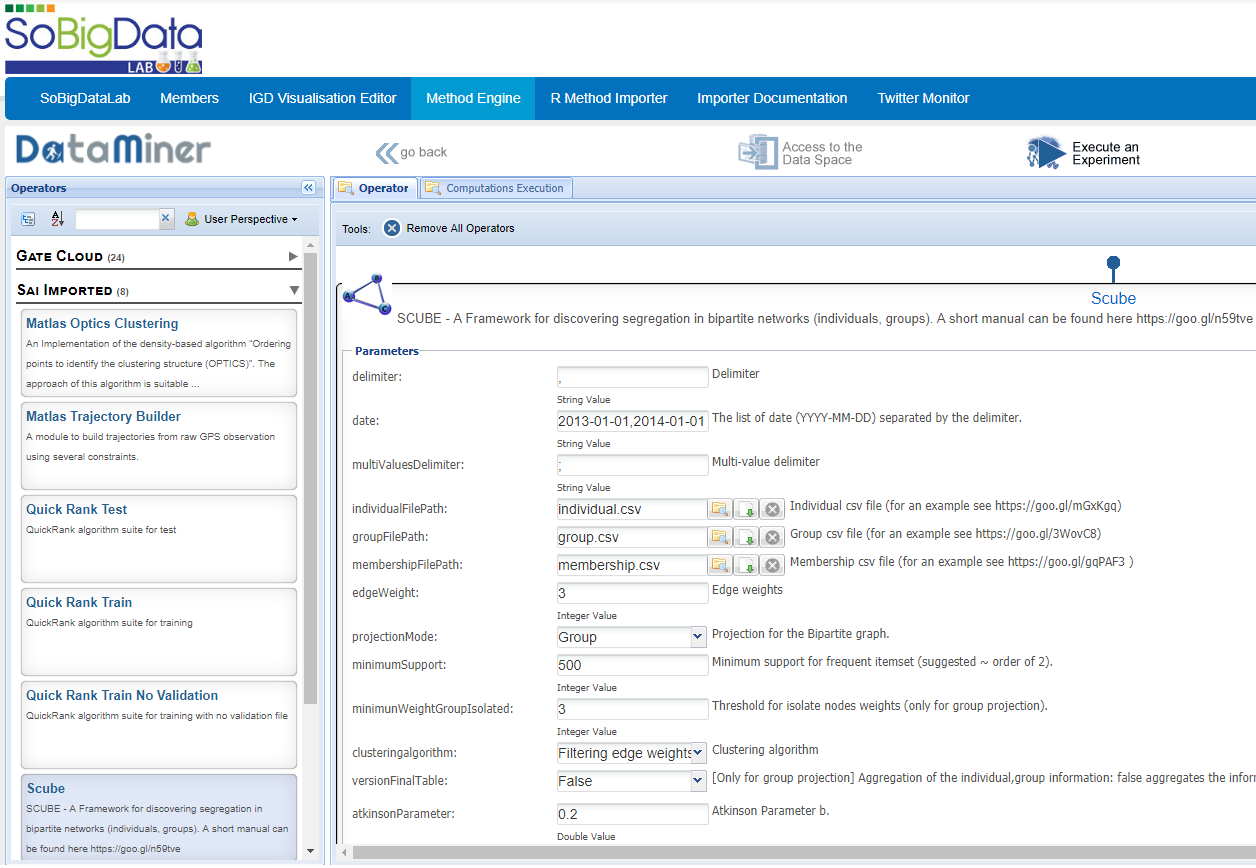}
	\caption{SCube standalone wizard (left) and SCube method at the SoBigData research infrastructure (right).}
	\label{fig:d4scienceWebInterfaceSnapshot}
	\vspace{-3mm}
\end{figure*}

\begin{figure}[t]
	\centering
	\includegraphics[scale=0.47]{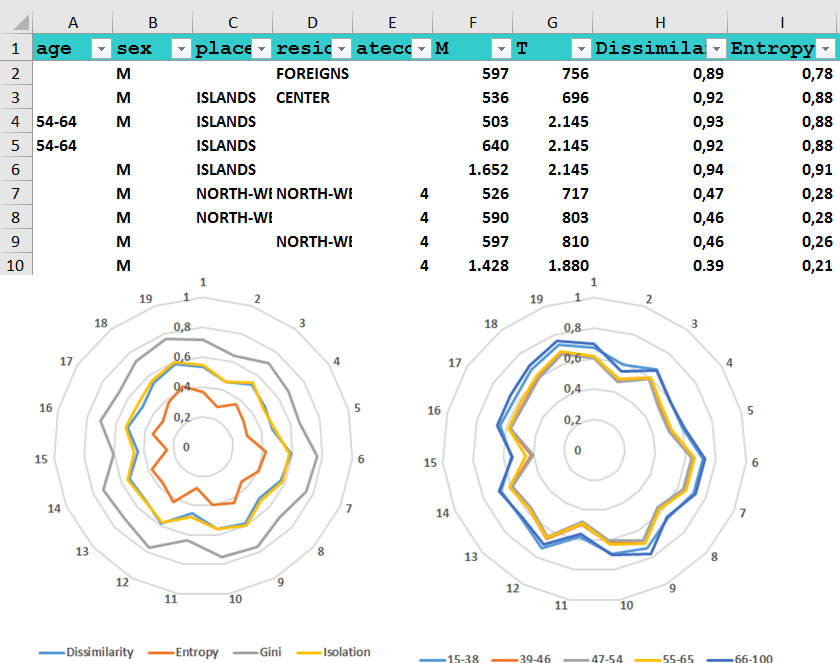}
	\caption{Top: sample multidimensional segregation cube. Bottom: radial plot of segregation indexes for directors in each of the 20 Italian company sectors.}
	\label{fig:excel}
	\vspace{-4mm}
\end{figure}

\textbf{Modules.} SCube consists of five software modules. \emph{GraphBuilder} projects the bipartite graph of individuals and groups into an unipartite attributed graph, where nodes are groups and an edge connect two
groups if they are related by at least one shared individual. 
In the case studies, nodes are companies, and edges connect companies that share at least
one director in their boards. Edges are weighted by the number of shared directors. \textit{GraphBuilder} outputs edges of the projection
(\texttt{\small edges}), and nodes that have zero degree (\texttt{\small isolated}). 
The \emph{GraphClustering} module computes then a clustering of nodes into organizational units (output file \texttt{\small nodeUnit}). Methods for clustering available in SCube include:
extraction of connected components (Breadth-First Search), removal of edges from the giant component with weight below a threshold and then
extraction of connected components (designed in \cite{Baroni2017jiis}), and an attributed graph clustering method for very large graphs (SToC algorithm \cite{Baroni2017asonam}). In our case studies, the result of \textit{GraphClustering} is a partitioning of companies into clusters based on connections among companies determined by shared directors -- which can be readily considered  a signal of relationships (business, personal, or other) between companies. Clusters represent
the organizational units needed for computing segregation indexes. \emph{TableBuilder} joins features of individuals with features of the companies
in an organizational unit. This yields a \texttt{\small finalTable} with a row per individual and organizational unit she belongs to. An example is
shown in Fig.~\ref{fig:KDDProcess} (left, bottom). This is the input for the \emph{SegregationDataCube} builder module, implementing the algorithm of \cite{Baroni2017jiis}. Notice that if the data under analysis contains already the assignment of individuals to units,~i.e., it is already in the form of \texttt{\small finalTable}, the pre-processing steps of bipartite projection and graph clustering do not need to be performed.
The \emph{Visualizer} module transforms the extended datacube in output of SegregationDataCube into a standard OOXML format that can be opened by Microsoft Excel, Libre Office, and other office productivity tools (see Fig.~\ref{fig:excel}).
Segregation data cube exploration can be easily interfaced with visualization tools, as in the map overlay in Fig.~\ref{fig:KDDProcess} (right).

\textbf{Process, Wizard, and GUI.} The whole process of segregation
discovery supported by SCube is shown in Fig.~\ref{fig:KDDProcess} (left, top). 
To facilitate the adoption of SCube by non-technical users, we have developed two interfaces (see Fig.~\ref{fig:d4scienceWebInterfaceSnapshot}).
The first one is a standalone wizard that guides the user throughout all the steps of the process, asking for
inputs and parameters when appropriate, and finish launching Microsoft Excel or Libre Office on the output file. Using popular desktop tools as GUI's makes the learning curve of approaching and effectively using SCube more manageable.
The second one is a cloud service offered by the SoBigDataLab freely accessible research infrastructure (\href{http://www.sobigdata.eu/access/virtual}{www.sobigdata.eu/access/virtual}), a web front-end comprising a catalogue of data, services, and virtual research environments for big data and social mining research.

\section{Demonstration Scenario}

The demonstration starts with a brief introduction on concepts and methods of segregation measurement \cite{massey1988dimensions} and segregation discovery \cite{Baroni2017jiis}. This provides the audience with the basic definitions for understanding the SCube functionalities. The  architecture of SCube is presented next. For interested participants, computational efficiency, algorithmic solutions,  and source code internal aspects are discussed. 
%
Then, two running case studies in the context of occupational segregation in the boards of company directors \cite{Aluncha2018} are introduced. They are based on a 2012 snaphost of  Italian companies (3.6M directors, 2.15M companies), and on a 20-year long dataset of Estonian companies (440K directors, 340K companies). Such anonymized datasets are the largest ever considered in the literature of segregation analysis. We summarize the data pre-processing activities to produce the inputs for~SCube.

The demonstration then proceeds by presenting three analysis scenarios based on input data of increasing complexity. In all scenarios, \texttt{\small gender}, \texttt{\small age}, and \texttt{\small birthplace} are used as segregation attributes. The first scenario considers tabular data, where company sector is used as organizational \texttt{\small unitID}, and it is intended to answer questions such as: how much are women segregated in company sectors? The second scenario considers attributed graph data, where nodes are directors, and edges connect two directors if they belong to a same company board. Here, the organizational units are determined through clustering over attributed graphs. This scenario can answer questions such as: how much are women segregated in communities of connected directors? Finally, the third scenario considers a bipartite attributed graph of directors and companies, as presented throughout the paper. An example of question it can answer is: how much are women segregated in communities of connected companies? For each scenario, 
the output of SCube is interactively explored using pivot tables and charts. The audience is guided to the discovery of a few actual cases of \textit{a-priori} unknown segregation contexts and to the understanding of which attributes contribute the most to segregation. Moreover, a cross-comparison of the Italian vs Estonian segregation findings will be discussed.


\section{Conclusion}

This demonstration illustrates the SCube tool for interactive exploration of social segregation indexes in large and complex data. The audience is made aware of social exclusion issues that can be hidden in data and of the indexes that measure segregation. Real case studies on scenarios of increasing complexity are discussed and explored. Efficiency issues and algorithmic solutions adopted for scaling to large datasets and graphs are detailed.


\medskip
{\textbf{Acknowledgements.}
This work is partially supported by the European H2020 Program under the funding scheme ``INFRAIA-1-2014-2015: Research Infrastructures'' grant agreement 654024 \emph{``SoBigData''} (\href{http://www.sobigdata.eu}{http://www.sobigdata.eu}).}

\bibliographystyle{abbrv}

\end{document}